\documentclass[
nofootinbib,
byrevtex,
showpacs,
showkeys
]{revtex4}
\usepackage[LY1]{fontenc}
\usepackage{amsmath,amssymb,amsxtra}
\usepackage[dvipdfm]{graphicx}
\usepackage{float}
\usepackage[dvipdfm,bookmarks=true,bookmarksnumbered=true,%
bookmarkstype=toc]{hyperref}
\AtBeginDvi{}
\makeatletter
\def\@pdfm@dest#1{%
  \Hy@SaveLastskip
  \@pdfm@mark{dest (#1) [@thispage /\@pdfview\space @xpos @ypos null]}%
  \Hy@RestoreLastskip
}
\makeatother
\newcommand{\bs}[1]{\boldsymbol{#1}}
\newcommand{\pa}{\partial}
\newcommand{\al}{\alpha}
\newcommand{\del}{\delta}

\begin{document}
\title{Evolution of Non-linear Fluctuations in Preheating after Inflation}
\author{Yasusada Nambu}
\email{nambu@gravity.phys.nagoya-u.ac.jp}
\author{Yohei Araki}
\email{araki@gravity.phys.nagoya-u.ac.jp}
\affiliation{Department of Physics, Graduate School of Science, Nagoya 
University, Chikusa, Nagoya 464-8602, Japan}
%
\date{November 25, 2005}
\begin{abstract}
 We investigate the evolution of  the non-linear long wavelength
 fluctuations during preheating after inflation. By using the separate
 universe approach, the temporal evolution of the power spectrum of the
 scalar fields and the curvature variable is obtained numerically.
  We found that the amplitude of the large scale fluctuations is
  suppressed after non-linear evolution during preheating.
\end{abstract}

\maketitle

\section{Introduction}
Preheating after inflation is a crucial stage in the early
universe. Fluctuations of matter fields and gravitational fields are
amplified due to  the parametric resonance caused by the coherent
oscillation of the inflaton
field\cite{TraschenJ:PRD42:1990,KofmanL:PRL73:1994,KodamaH:PTP96:1996,
NambuY:1997,TaruyaA:1998}.
One of the important question on preheating is how the long wavelength
metric fluctuation is amplified by the parametric resonance. The
linear analysis shows that non-adiabatic modes of long wave
fluctuations are amplified and grow during preheating until the
non-linear effect by the second order fluctuations becomes
dominant. When the non-linear effect cannot be neglected,  the
backreaction of the fluctuations on the evolution of background
quantities  becomes dominant and the amplification of long wave modes
stops \cite{BassetB:PRD62:2000,ZibinJP:PRD63:2001}. To obtain complete
understanding of the evolution of inhomogeneities during  preheating,
we have to investigate non-linear evolution of fluctuations .

The separate universe approach\cite{WandsD:2000} is an appropriate method
to treat  the non-linear dynamics of long wavelength fluctuations during
preheating. This approach neglects fluctuations of which wavelength is
smaller than the Hubble horizon scale. For each spatial point, the
basic equation of the separate universe reduces to that of the
Friedmann equation for a homogeneous and isotropic flat
universe. However, all dynamical variables include non-linear
inhomogeneities of which wavelength is larger than the Hubble horizon
scale. As the separate universe approach includes all order long
wavelength non-linear gravitational fluctuations, we can apply this
method to investigate the backreaction effect on  evolution of a
Friedmann-Robertson-Walker(FRW) universe\cite{NambuY:PRD71:2005}.

Tanaka and Basset\cite{TanakaT:2003} investigated preheating using the
separate universe approach. They found that initial small fluctuations
are amplified and evolve to random spatial distribution. In the
separate universe, dynamical variables at each spatial point evolve
independently. Hence, in the early stage of evolution, fluctuations
keep their initial spatial distribution  and only the amplitude of
fluctuations  grows by the parametric resonance. When the amplitude of
the massless field grows to be comparable to that of the inflaton
field, the non-linear interaction between scalar fields becomes to be
dominant and the system enters a chaotic regime. Then,  the field
variables at each spatial point behave as independent random
variables. At this stage, the power spectrum of fluctuations is  same
as that of random white noise.

In this paper, we concentrate on the evolution of the power spectrum
of fluctuations during preheating especially on curvature fluctuation
that gives an impact on formation of large scale structure. By performing the numerical
simulation of preheating based on the separate universe approach, we aim to
understand the feature of evolution of long wavelength non-linear
fluctuation during preheating.

The plan of paper is as follows. In Sec.~2, we review the separate
universe approach. In Sec.~3, we analytically estimate the evolution
of the power spectrum of scalar fields and a curvature variable. We
present our numerical results in Sec.~4 and Sec.~5 is devoted to
summary and conclusion. We use units in which $c=\hbar=8\pi G=1$
throughout the paper.

\section{Short review of the separate universe approach}
The separate universe approach\cite{WandsD:2000} is the lowest order spatial gradient
expansion (GE)\cite{SalopekDS:1992} that is obtained by neglecting  all terms containing
the second order spatial derivative terms in the Einstein equation.
All dynamical variables keep the spatial dependence that
represents large scale inhomogeneities whose wavelength is larger than
the Hubble horizon scale. In this approach, we assume the
following form of the metric :
\begin{equation}
  \label{eq:metric}
  ds^2=-N^2 dt^2+e^{2\al}d\bs{x}^2,
\end{equation}
where $N$ is a lapse function and $\al$ is the logarithm of the scale
factor of the universe. 
For the system with two scalar fields $\phi_i~(i=1,2)$,
the Einstein equation becomes
\begin{align}
&3H^2=\frac{1}{2}\sum_j\left(\frac{\dot\phi_j}{N}\right)^2+V(\phi_1,\phi_2),
\label{eq:HC}\\
&\pa_i H=-\frac{1}{2N}\sum_j\dot\phi_j\pa_i\phi_j,
\label{eq:MC} \\
&\frac{\dot\al}{N}=H,\quad
\frac{\dot H}{N}=
-\frac{1}{2}\sum_j\left(\frac{\dot\phi_j}{N}\right)^2, \label{eq:evo-geometry}\\
&\frac{1}{N}\left(\frac{\dot\phi_j}{N}\right)\spdot
+3H\left(\frac{\dot\phi_j}{N}\right)+\frac{\pa V}{\pa\phi_j}=0,
\label{eq:evo-scalar}
\end{align}
where Eq.~\eqref{eq:HC} is the Hamiltonian constraint, Eq.~\eqref{eq:MC} is
the momentum constraint, Eq.~\eqref{eq:evo-geometry} and
Eq.~\eqref{eq:evo-scalar} are evolution equations for the scalar factor
and scalar fields.

To understand how the spatial inhomogeneity is included in dynamical
variables, it is convenient to use the Hamilton-Jacobi
method\cite{SalopekDS:1992}. By introducing the Hubble function
$H=H(\phi_1,\phi_2)$ that is a function of scalar fields, the above set of
equations can be written
\begin{align}
  &3H^2=2\sum_j\left(\frac{\pa H}{\pa\phi_j}\right)^2+V(\phi_1,\phi_2), 
 \label{eq:HJ-HC}\\
  &\pa_i H=\sum_j\frac{\pa H}{\pa\phi_j}\pa_i\phi_j,
  \label{eq:HJ-MC}\\
  &\frac{\dot\al}{N}=H, \label{eq:HJ-evo-geomerty}\\
  &\frac{\dot\phi_j}{N}=-2\frac{\pa H}{\pa\phi_j}.
   \label{eq:HJ-evo-scalar}
\end{align}
The solution of the Hamilton-Jacobi equation \eqref{eq:HJ-HC} contains
two constants of integration $d_1(\bs{x}), d_2(\bs{x})$.  If these
constants do not have spatial dependence\footnote{The
  constants $d_j$ can have spatial dependence if they satisfy
  $H_{,\phi_1}\pa_i d_1+H_{,\phi_2}\pa_i d_2=0$. However, the
  inhomogeneity caused by $d_i$ is related to the decaying mode of 
  perturbations and sub-dominant. Hence we ignore this type of
  inhomogeneity here.}, the Hubble function
$H=H(\phi_1(\bs{x}),\phi_2(\bs{x});d_1, d_2)$ satisfies the momentum
constraint  \eqref{eq:HJ-MC}.
By differentiating $H$ with respect to $d_j$, we have other
integration of constants $c(\bs{x}), f(\bs{x})$:
\begin{align}
  &e^{3\al}\frac{\pa H}{\pa d_1}\equiv e^{-3c(\bs{x})}, \\
  &e^{3\al}\frac{\pa H}{\pa d_2}\equiv e^{-3c(\bs{x})}f(\bs{x}). 
\end{align}
By solving these equations with respect to $\phi_j$, we can
obtain the scale factor dependence of the scalar fields:
\begin{align}
  &\phi_1=\phi_1(\al+c(\bs{x}); f(\bs{x}),d_1,d_2), \notag\\
  &\phi_2=\phi_2(\al+c(\bs{x}); f(\bs{x}),d_1,d_2). \label{eq:phi-sol}
\end{align}
The gauge invariant variable that reduces to the spatial derivative of
the scalar field on the zero curvature gauge is defined by\cite{RigopoulosGI:2003}
\begin{equation}
  Q_i^{(j)}=\pa_i\phi_j-\phi_{j,\al}\pa_i\al,
\end{equation}
and these variables are  non-linear generalization of Mukhanov's gauge
invariant variable in the linear perturbation.  The scalar field
equation \eqref{eq:evo-scalar} in the zero curvature slice
$\pa_i\al=0$ is
$$
\phi_{j,\al\al}+\left(3+\frac{H_{,\al}}{H}\right)\phi_{j,\al}
+\frac{V_{,\phi_j}}{H^2}=0,
$$
and by taking the spatial derivative of this equation, it is possible
to show that the variable $\bs{Q}_i=(Q_i^{(1)},Q_i^{(2)})$ satisfies
the following evolution equation:
\begin{align}
&\bs{Q}_{i,\al\al}+\left(3+\frac{H_{,\al}}{H}\right)\bs{Q}_{i,\al}
+\bs{M}\bs{Q}_i=0, \label{eq:Mukhanov}\\
&; (\bs{M})_{ij}=\frac{V_{\phi_i\phi_j}}{H^2}
-\frac{e^{-3\al}}{H}\left(e^{3\al}H\phi_{i,\al}\phi_{j,\al}\right)_{,\al},\notag
\end{align}
that  has the exactly same form as in the linear perturbation
case. However, the coefficients of the equations are spatial dependent
and this equation describes non-linear evolution of large scale
inhomogeneities. 
The non-linear generalization of the gauge invariant variable in the
linear perturbation that
reduces to the spatial derivative of the three curvature on the
comoving slice is defined by
\begin{equation}
  \mathcal{R}_i=\pa_i\al-\frac{NH}{\dot\rho_\text{tot}}\pa_i\rho_\text{tot}
  =\pa_i\al-\frac{\sum_j\phi_{j,\al}\pa_i\phi_j}{\sum_j\phi_{j,\al}^2}
  =\pa_i\al-\frac{H}{\dot H}\pa_iH, \label{eq:Ri}
\end{equation}
and this quantity can be written  using $Q_i$:
\begin{equation}
  \mathcal{R}_i=-\frac{\sum_j\phi_{j,\al}Q_i^{(j)}}{\sum_j\phi_{j,\al}^2}.
\end{equation}
By substituting the solution \eqref{eq:phi-sol}, the gauge invariant quantity
$\bs{Q}$ becomes
\begin{equation}
\label{eq:sol-scalar}
\bs{Q}_i=
\begin{pmatrix}
\phi_{1,\al} & \phi_{1,f} \\
\phi_{2,\al} & \phi_{2,f}
\end{pmatrix}
\begin{pmatrix}
\pa_ic(\bs{x}) \\ \pa_i f(\bs{x})
\end{pmatrix}
-\begin{pmatrix}\phi_{1,\al}\\ \phi_{2,\al}\end{pmatrix}\pa_i\al,
\end{equation}
and the gauge invariant variable $\mathcal{R}_i$ becomes
\begin{equation}
  \mathcal{R}_i=\pa_i\al-\pa_ic
  +\frac{\sum_j\phi_{j,\al}\phi_{j,f}}{\sum_j\phi_{j,\al}^2}\pa_if.
\end{equation}
If we take the zero curvature slice $\pa_i\al=0$,  
\begin{equation}
\label{eq:Ri}
\mathcal{R}_i=-\pa_ic
  +\frac{\sum_j\phi_{j,\al}\phi_{j,f}}{\sum_j\phi_{j,\al}^2}\pa_i f.
\end{equation}
From this formula, we can observe that the curvature fluctuation
caused by $c(\bs{x})$ does not evolve. This mode corresponds to the
adiabatic mode in linear perturbation. The curvature fluctuation
caused by $f(\bs{x})$ corresponds to the non-adiabatic  mode and the
evolution of the curvature fluctuation is caused by the non-adiabatic
mode of scalar fields fluctuations.

\section{Evolution of inhomogeneities in preheating} 
In this section, we consider a specific inflationary
model with the inflaton field $\phi$ and the massless scalar field
$\chi$. The potential is assumed to be 
\begin{equation}
  \label{eq:potential}
  V(\phi, \chi)=\frac{\lambda}{4}\phi^4+\frac{g^2}{2}\phi^2\chi^2.
\end{equation}
For $\phi\lesssim m_{pl}$, the inflaton field oscillate coherently
about $\phi=0$ and the evolution of the $\chi$ field gets the effect
of the parametric resonance. For  $g^2/\lambda=2$ case, the longest wavelength
  mode $k=0$ is included in a strong resonance band and the
fluctuation grows exponentially in time with Floquet index
$\mu\approx 0.238$. Thus, the super-horizon mode is amplified by the
parametric resonance. As we are interested in the non-linear evolution
of fluctuations, we consider $g^2/\lambda=2$ case in which the growth
rate for $k=0$ fluctuation is the largest.

\subsection{Analytic approximation}
In preheating model defined by \eqref{eq:potential},  the time averaged
equation of state of the inflaton  fields is  same as that of
radiation and the evolution of the scale factor is given by
\begin{equation}
  e^{\al}\approx \left(\frac{t}{t_0}+1\right)^{1/2},
\end{equation}
where we choose $\al=0$ at $t=0$. By introducing the conformal time
$\eta=\int dte^{-\al}$ and the conformal variables
$\tilde\phi=e^{\al}\phi, \tilde\chi=e^{\al}\chi$, the evolution
equations of scalar fields become
\begin{align}
  &\tilde\phi''+\lambda\tilde\phi^3+g^2\tilde\phi\tilde\chi^2\approx 0,
  \notag \\
  &\tilde\chi''+g^2\tilde\phi^2\tilde\chi\approx 0,\label{eq:flat-scalar} 
\end{align}
where $'=d/d\eta$. The curvature variable \eqref{eq:Ri} becomes
\begin{equation}
  \mathcal{R}_i\approx
  -\frac{1}{4}\left(\frac{\eta_0}{\eta}\right)^2\left[
    \tilde\phi'\pa_i\tilde\phi+\tilde\chi'\pa_i\tilde\chi
+O\left(\frac{1}{\eta}\right)\right] 
\end{equation}
The solution of the separate universe can be written as
\begin{equation}
  \tilde\phi=\tilde\phi(\eta,
  f(\bs{x})),\quad\tilde\chi=\tilde\chi(\eta, f(\bs{x})),
\end{equation}
where $f(\bs{x})$ is a spatially dependent constant of integration. 
 We obtain perturbative solution by assuming that the amplitude of
 $\tilde\chi$ is small:
\begin{equation}
  \tilde\phi=\tilde\phi_0+\tilde\phi_1+\cdots,\qquad
  \tilde\chi=\tilde\chi_1+\tilde\chi_2+\cdots.
\end{equation}
Then the equations for each order of the perturbation become
\begin{align}
  &\tilde\phi_0''+\lambda\tilde\phi_0^3=0, \\
  &\tilde\chi_1''+g^2\tilde\phi_0^2\tilde\chi_1=0,\quad\tilde\phi_1=0,\\
  &\tilde\phi_2''+3\lambda\tilde\phi_0^2\tilde\phi_2=-g^2\tilde\phi_0\tilde\chi_1^2,
  \quad\tilde\chi_2=0, \\
  & \tilde\chi_3''+g^2\tilde\phi_0^2\tilde\chi_3=-2g^2\tilde\phi_0\tilde\phi_2\tilde\chi_1.
\end{align}
 The solution of the background inflaton field $\tilde\phi_0$ is given
 by
\begin{equation}
  \tilde\phi_0=c\,\mathrm{cn}(\lambda^{1/2}c\eta; 1/2)
\end{equation}
where $\mathrm{cn}$ is the Jacobi elliptic cosine, $c$ is the initial
value of $\tilde\phi_0$. We approximate the elliptic function as
follows\cite{PBGreeneLKofmanAL:PRD56:1997}:
\begin{align}
  &\tilde\phi_0\approx c\,\cos(\tau),\quad
  \tilde\phi_0^2\approx c^2\left[F_0+F_1\cos(2\tau)\right],\\
  &\tau=\frac{2\pi}{T}\lambda^{1/2}c\,\eta,\quad
  T=7.416,~F_0=0.457,~F_1=0.4973. \notag
\end{align}
The equation of $\tilde\chi_1$ becomes the Mathieu equation and
$\tilde\chi_1$ grows by the parametric resonance:
\begin{equation}
  \tilde\chi_1=fe^{\mu\tau}\cos\tau,\quad \mu\approx 0.12\frac{g^2}{\lambda}
\end{equation}
where $f$ is a spatially dependent constant of integration which
specifies the initial amplitude of $\tilde\chi$. The
solution up to the third order becomes
\begin{align}
  &\tilde\phi=\left[c
  -c_1\left(\frac{g^2}{\lambda}\right)\frac{f^2}{
    c}e^{2\mu\tau}\right]\cos(\tau) +\text{($\sin$ part $+$ higher frequency part)}, \\
  &\tilde\chi=\left[f
    e^{\mu\tau}-c_2\left(\frac{g^2}{\lambda}\right)
    \frac{f^3}{c^2}e^{3\mu\tau}\right]\cos(\tau)+\text{($\sin$ part
    $+$ higher frequency part)},\\
  &\tau\approx
  \lambda^{1/2}c\eta, \notag
\end{align}
where $c_1, c_2$ are $O(1)$ numerical factors.
 At the time given by
\begin{equation}
  \eta_*\sim\frac{1}{\mu\lambda^{1/2}c}
\ln\left(\frac{c}{f}\sqrt{\frac{\lambda}{g^2}}\right), \label{eq:time-scale}
\end{equation}
the amplitude of the higher order perturbation terms grow to be
comparable to that of the lower order quantities. At this time,
fluctuations of $\tilde\chi$ field changes the evolution of
$\tilde\phi$ field through back reaction effect and  the exponential
growth of fluctuation of $\tilde\chi$-field is shut off. After this
time, non-linearities become dominant.

As the  Eq.~\eqref{eq:flat-scalar} is a chaotic system, the behavior
of the scalar fields becomes chaotic after this time. The time scale
\eqref{eq:time-scale} depends on the initial value $f(\bs{x})$ of
$\tilde\chi$-field that is the spatial dependent function. The
deviation of the field of neibouring two  spatial points grows
exponentially for $\eta<\eta_*$. After $\eta>\eta_*$, as the system
is chaotic, the dependence of the initial condition is
completely randomized and the spatial distribution of $\tilde\chi$
becomes that of white noise\cite{TanakaT:2003}.

To evaluate the evolution of the power spectrum of $\tilde\chi$, we
assume the function $f(\bs{x})$ that specifies the initial
distribution of $\chi$ obeys the random Gaussian
statistics:
\begin{equation}
   f(\bs{x})=\sum_{\bs{k}}f_{\bs{k}}\,e^{i\bs{k}\cdot\bs{x}},\quad
   \langle
   f_{\bs{k}_1}f^*_{\bs{k}_2}\rangle=
f_0^2\left(\frac{k}{k_c}\right)^n\del_{\bs{k_1}\bs{k_2}}
\quad f_{\bs{k}}=f^{*}_{-\bs{k}}, 
\end{equation}
where $k_c$ is a cut off of wave number and $f_0$ defines the
amplitude of the power at $k=k_c$.  The actual initial power spectrum
for the $\chi$-field expected to be produced during inflation in this
model is investigated by Zibin \textit{et
  al}.\cite{ZibinJP:PRD63:2001} Owing to their result, the initial
spectrum of $\chi$-field  for large scales is given by $n=-3$ and scale
invariant. In this paper, as we are interested in feature of
non-linear evolutions of fluctuations, we does not fix the value of
power index of $\chi$-field. The power spectrum of $\tilde\chi$ field
is given by
\begin{equation}
  \mathcal{P}_{\tilde\chi}=(k/k_c)^3\langle|\tilde\chi_{\bs{k}}|^2\rangle
 \approx C_1e^{2\mu\tau}\left(\frac{k}{k_c}\right)^{n+3}
 +C_2\frac{g^2}{\lambda c^2}e^{4\mu\tau}\left(\frac{k}{k_c}\right)^{2n+6},
\end{equation}
where $C_1, C_2$ are $O(1)$ numerical factors. At $\eta\sim \eta_*$,
the amplitude of the fluctuation with the wave number
\begin{equation}
  \left(\frac{k_*}{k_c}\right)\sim
  \left(\frac{\lambda c^2}{g^2}\right)^{\frac{1}{n+3}}
\exp\left(-\frac{2\mu\lambda^{1/2}c\eta_*}{n+3}\right).
\end{equation}
becomes $O(1)$ and non-linear. Once the fluctuation with the wave
number $k_*$ becomes non-linear, as the system is chaotic, the
coherence beyond the scale $1/k_*$ is lost and the fluctuation with
the wave number $k<k_*$ is randomized  and behaves as  white
noise. Thus, the power spectrum for $k<k_*$ becomes that of white noise
$\propto k^3$. The evolution of the shape of the power spectrum
depends on the value of the initial power index $n$.

For $n<-3$, as the power of  smaller $k$ mode is larger, the
smaller $k$ mode (large scale)  goes to be non-linear first. At
$\eta\sim\eta_*$, fluctuation with the wave number $k<k_*$ reaches 
non-linear  and the power spectrum becomes
$\mathcal{P}_{\tilde\chi}\sim k^3$. At this time, fluctuation with the
wave number $k>k_*$ stays linear and the spectrum keeps its initial
shape $\mathcal{P}_{\tilde\chi}\sim k^{n+3}$. Thus, the shape of the
power spectrum at $\eta\sim\eta_*$ is given by
\begin{equation}
 \mathcal{P}_{\tilde\chi}(k)\sim
 \begin{cases}
   k^3\quad& (k<k_*) \\
   k^{n+3}\quad & (k>k_*)
 \end{cases}
\end{equation}
 For $n>-3$, large $k$ modes (small scale) goes to non-linear
 first. Once the smallest scale of the system becomes non-linear at
 $\eta\sim\eta_*$, the coherence of larger scale fluctuation is
 lost and the spectrum  for all wave number changes to be
 that of the white noise:
\begin{equation}
 \mathcal{P}_{\tilde\chi}(k)\sim
   k^3\quad (k<k_*) 
\end{equation}

The power spectrum of the curvature variable $\mathcal{R}_i$ is given by
\begin{equation}
  \mathcal{P}_{\mathcal{R}_i}(k)\approx
  C_3\frac{e^{4\mu\tau}}{\tau^4}\left(\frac{k}{k_c}\right)^{n+5}
  +C_4\frac{g^2}{\lambda\mu^2c^2}\frac{e^{6\mu\tau}}{\tau^4}
\left(\frac{k}{k_c}\right)^{2n+8}
\end{equation}
where $C_3, C_4$ are $O(1)$ numerical factors. At $\eta\sim\eta_*$, the
power spectrum for $n<-3$ is
\begin{equation}
 \mathcal{P}_{\mathcal{R}_i}(k)\sim
 \begin{cases}
   k^3\quad& (k<k_*) \\
   k^{n+5}\quad & (k>k_*)
 \end{cases}
\end{equation}
For $n>-3$, the spectrum at $\eta\sim\eta_*$ is
\begin{equation}
 \mathcal{P}_{\mathcal{R}_i}(k)\sim
   k^3\quad (k<k_*) 
\end{equation}
We summarize the result of the analytic estimation of the power
spectrum in TABLE \ref{tbl:a}. 
\begin{table}[H]
\centering
 \begin{tabular}{|l|c|c|} \hline
 \multicolumn{3}{|c|}{$n<-3$} \\ \hline
 & $\eta<\eta_*$ & $\eta>\eta_*$ \\ \hline
 $\mathcal{P}_\chi$ & $k^{n+3}$ & $k^3~(k<k_*),\quad k^{n+3}~(k>k_*)$\\ \hline
 $\mathcal{P}_{\mathcal{R}_i}$ & $k^{n+5}$ &
 $k^3~(k<k_*),\quad k^{n+5}~(k>k_*)$ \\ \hline
 \end{tabular}
\quad
  \begin{tabular}{|l|c|c|} \hline
 \multicolumn{3}{|c|}{$n>-3$} \\ \hline
 & $\eta<\eta_*$ & $\eta>\eta_*$ \\ \hline
 $\mathcal{P}_\chi$ & $k^{n+3}$ &  $k^3 $\\ \hline
 $\mathcal{P}_{\mathcal{R}_i}$ & $k^{n+5}$ & $k^3$ \\ \hline
  \end{tabular}
 \caption{\label{tbl:a}Power spectrums of $\chi$ and $\mathcal{R}_i$ for
   $3$-dimensional  space.  The initial $\chi$ is scale invariant for $n=-3$.}
\end{table}
To compare the numerical calculation performed in the following sections, we
present the estimation of the power spectrum for 1-dimensional
case (TABLE \ref{tbl:b}). The initial spectrums are assumed to be
\begin{equation}
 \mathcal{P}_{\chi}\propto k^{n+1},\quad
 \mathcal{P}_{\mathcal{R}_i}\propto k^{n+3}.
\end{equation}
\begin{table}[H]
  \centering
  \begin{tabular}{|l|c|c|} \hline
 \multicolumn{3}{|c|}{$n<-1$} \\ \hline
 & $\eta<\eta_*$ & $\eta>\eta_*$ \\ \hline
 $\mathcal{P}_\chi$ & $k^{n+1}$ & $ k~(k<k_*),\quad k^{n+1}~(k>k_*) $\\ \hline
 $\mathcal{P}_{\mathcal{R}_i}$ & $k^{n+3}$ &
 $k~(k<k_*),\quad k^{n+3}~(k>k_*)$ \\ \hline
  \end{tabular}
  \qquad
  \begin{tabular}{|l|c|c|} \hline
 \multicolumn{3}{|c|}{$n>-1$} \\ \hline
 & $\eta<\eta_*$ & $\eta>\eta_*$ \\ \hline
 $\mathcal{P}_\chi$ & $k^{n+1}$ & $k$\\ \hline
 $\mathcal{P}_{\mathcal{R}_i}$ & $k^{n+3}$ &
 $k$ \\ \hline
  \end{tabular}
  \caption{ \label{tbl:b}Power spectrums of $\chi$ and $\mathcal{R}_i$ for
    $1$-dimensional space. The initial $\chi$ is scale
    invariant for $n=-1$.}
\end{table}

\subsection{Numerical simulation of preheating}
To confirm the analytic estimation of the evolution of the power spectrums, we perform
1-dimensional lattice simulation of the separate universe.

By introducing the momentum variables $P_{\phi}=\dot\phi/N,
P_{\chi}=\dot\chi/N$, our basic equations of the separate universe are
\begin{align}
  &3H^2=\frac{1}{2}(P_{\phi}^2+P_{\chi}^2)+V(\phi,\chi), \label{eq:hc} \\
  &\pa_iH=-\frac{1}{2}(P_{\phi}\pa_i\phi+P_{\chi}\pa_i\chi), \label{eq:mc}\\
  &\frac{\dot\al}{N}=H,\quad \frac{\dot
    H}{N}=-\frac{1}{2}(P_{\phi}^2+P_{\chi}^2), \\
  &\frac{\dot\phi}{N}=P_{\phi},\quad \frac{\dot
    P_{\phi}}{N}=-3HP_{\phi}-\frac{\pa V}{\pa\phi}, \\
  &\frac{\dot\chi}{N}=P_{\chi},\quad \frac{\dot
    P_{\chi}}{N}=-3HP_{\chi}-\frac{\pa V}{\pa\chi}.
\end{align}
As the initial condition, we must prepare variables $(H, P_\phi,
P_\chi, \phi, \chi)$ that satisfy the
Hamiltonian constraint \eqref{eq:hc} and the momentum constraint
\eqref{eq:mc}. We adopt the following form of the initial condition :
\begin{equation}
  H=h_0,\quad P_{\chi}=0,\quad \chi=\chi_0({x}),\quad\phi=\phi_0,
\end{equation}
where $h_0$ and $\phi_0$ are spatially homogeneous constants and
$\chi_0({x})$ is the spatially dependent function that specifies
the initial inhomogeneity of the $\chi$-field. We prepare $\chi_0(x)$
as a random Gaussian field:
\begin{equation}
  \chi_0(x)=\sum_{k>0}\hat{A}_k\cos(kx+\hat{\theta}_k)
\end{equation}
where $\hat{\theta}_k$ is a uniform random number in $[0,2\pi]$ and
$\hat{A}_k$ is a random number whose probability distribution is
given by the Rayleigh distribution
\begin{equation}
  P(A_k)\propto A_k\exp\left(-\frac{A_k^2}{\chi_0(k/k_c)^n}\right)
\end{equation}
where $n$ is the power index of the field. We choose  $k_c=\pi N_{\text{grid}}/L$
where $L$ is the size of the calculating region and $N_{\text{grid}}$ is number of
grids. This value is  the smallest wave number that corresponds to the
size of the system.   The  value of the momentum  of the inflaton
field is determined by the Hamiltonian constraint \eqref{eq:hc}:
\begin{equation}
  P_{\phi}=\left[6h_0^2-2V(\phi_0, \chi_0({x}))\right]^{1/2}. 
\end{equation}
Once the initial values that satisfy the Hamiltonian constraint and
the momentum constraint, these constraints are preserved
during temporal evolution.

We then must specify the time slicing condition to evolve the system. The
simplest one is the synchronous slicing $N=1$. In this slice, the
evolution equations become
\begin{align}
  &\dot\al=H,\quad \dot
    H=-\frac{1}{2}(P_{\phi}^2+P_{\chi}^2),  \label{eq:sync1}\\
  &\dot\phi=P_{\phi},\quad \dot
    P_{\phi}=-3HP_{\phi}-\lambda\phi^3-g^2\phi\chi^2,  \label{eq:sync2}\\
  &\dot\chi=P_{\chi},\quad \dot
    P_{\chi}=-3HP_{\chi}-g^2\phi^2\chi. \label{eq:sync3}
\end{align}
One of the other choice  is the comoving slice. In this slice, the
Hubble parameter $H$ is treated as a time parameter ($t=H$) and the evolution
equations become
\begin{align}
  &\frac{1}{N}\frac{\pa\al}{\pa
    H}=H,\quad\frac{1}{N}=-\frac{1}{2}(P_\phi^2+P_\chi^2), \\
  &\frac{1}{N}\frac{\pa\phi}{\pa H}=P_\phi,\quad
    \frac{1}{N}\frac{\pa P_{\phi}}{\pa H}=-3HP_{\phi}-\lambda\phi^3-g^2\phi\chi^2,  \label{eq:sync2}\\
  &\frac{1}{N}\frac{\pa\chi}{\pa H}=P_{\chi},\quad
    \frac{1}{N}\frac{\pa P_{\chi}}{\pa H}=-3HP_{\chi}-g^2\phi^2\chi.
\end{align}
The second equation of Eq.~(57) determines the lapse function $N$. In
our numerical calculation, we adopt the synchronous slicing $N=1$ for
the sake of the technical simplicity.
We also assume that the $\al$ is homogeneous at initial time and we
have only non-adiabatic mode of curvature fluctuation initially.

We solve Eqs~\eqref{eq:sync1}-\eqref{eq:sync3} using the 4th order Runge-Kutta method.

\section{numerical results}
The spatial grid number used is $2^{14}=16384$ and the power spectrums are obtained 
by taking ensemble average of 100 different calculations for different
initial distribution of $\chi$ field with the same spectral index. We used the
parameter $\lambda=0.1, g^2=0.2, \phi_0=0.5, \chi_0=10^{-5},h_0=0.03$.

\subsection{flat space case}
We first performed the numerical calculation of  the scalar fields
system in flat space \eqref{eq:flat-scalar} to investigate the evolution
of non-linear fluctuation without gravitational effect. We obtained
the evolution of the power spectrum of $\tilde\chi$-field numerically.

\begin{figure}[H]
  \centering
  \includegraphics[width=0.4\linewidth,clip]{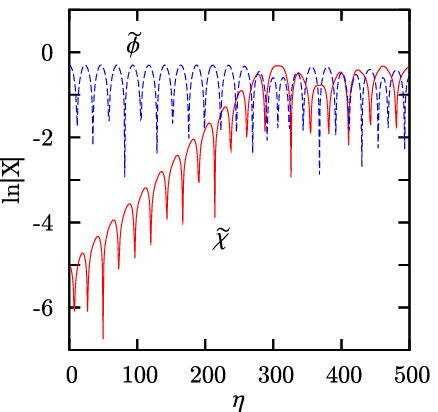}%
  \includegraphics[width=0.4\linewidth,clip]{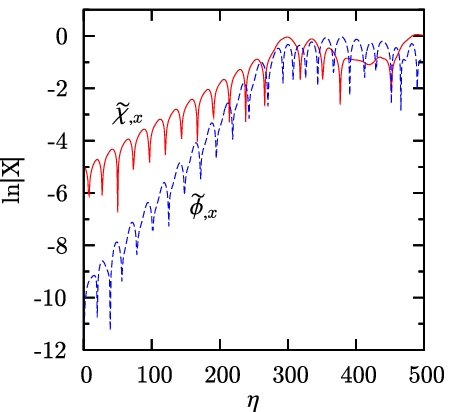}
 \caption{The left panel shows the evolution of the scalar fields at a
   specific spatial point. The right panel shows the evolution of the spatial
   derivative $\tilde\phi_{,x}$ and $\tilde\chi_{,x}$.}
  \label{fig:flat-field}
\end{figure}

The local evolution of the inflaton field and the
massless field is shown in Fig.~\ref{fig:flat-field}. The spatial derivative of
the fields is calculated for two different initial values
$\tilde\chi_0=0.5\times 10^{-5}, 10^{-5}$.
\noindent
In this calculation, the time scale given by
Eq.~\eqref{eq:time-scale} at which non-linear effect becomes
significant, is
$$
 \eta_*\sim 280.
$$
For $\eta\lesssim\eta_*$, the amplitude of $\tilde\chi$ field grows
exponentially in time by the parametric amplification while the
amplitude of $\tilde\phi$ field stays constant. In accordance with
this behavior, the spatial derivative of $\tilde\chi$ and $\tilde\phi$
grows exponentially in time. The growth rate of $\tilde\phi_{,x}$ is
greater than that of $\tilde\chi_{,x}$ and at $\eta\sim\eta_*$,
$\tilde\phi_{,x}\sim\tilde\chi_{,x}$. After $\eta\sim\eta_*$, the
system enters non-linear regime and the grow of $\tilde\chi,
\tilde\phi_{,x}, \tilde\chi_{,x}$ stops.

 We next investigate the evolution of the
power spectrum of $\tilde\chi$ for different initial power indices. 

\paragraph{$n=-3$ case }

\begin{figure}[H]
  \centering
\includegraphics[width=0.22\linewidth,clip]{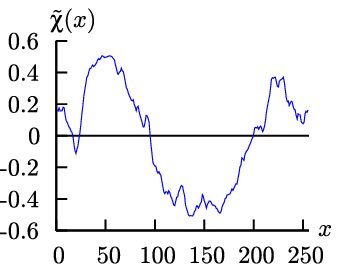}\hspace{0.5cm}
\includegraphics[width=0.22\linewidth,clip]{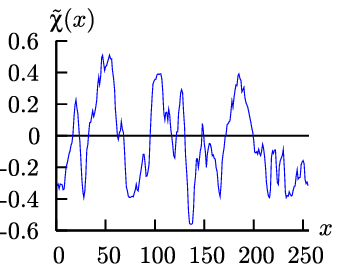}\hspace{0.5cm}
\includegraphics[width=0.22\linewidth,clip]{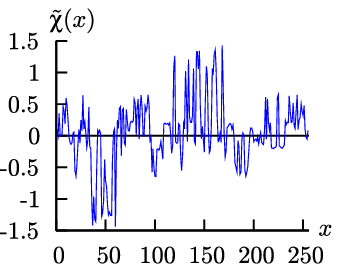}\hspace{0.5cm}
\includegraphics[width=0.22\linewidth,clip]{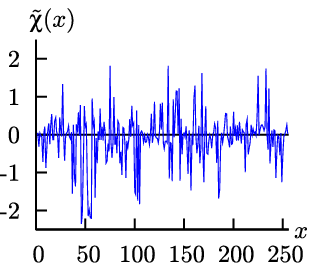}
  \caption{Evolution of the spatial distribution of $\tilde\chi$ field
    for the initial power index $n+1=-2$. From the left to the right
    panel, $\eta=0, 280, 310, 400$.}
\label{fig:pat1}
\end{figure}

\begin{figure}[H]
  \centering
\includegraphics[width=0.4\linewidth,clip]{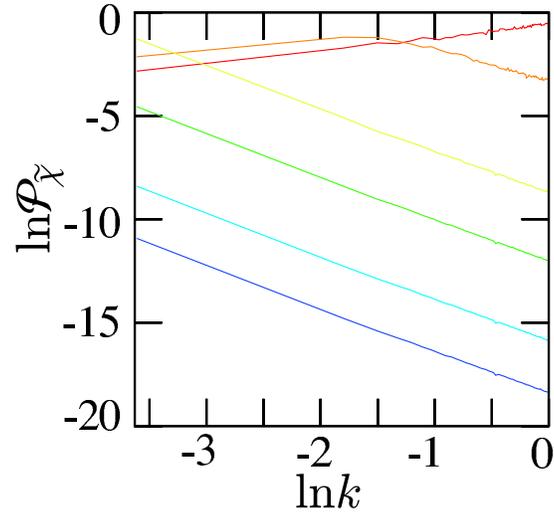}
  \caption{Evolution of the power spectrum  of $\tilde\chi$ field for the initial
    power index $n+1=-2$. The spectrums  at time $\eta=0\text{(blue)},
    100\text{(cyan)}, 200\text{(green)},
    300\text{(yellow)}, 400\text{(orange)},500\text{(red)}$ are shown.}
  \label{fig:spec1}
\end{figure}

Fig.~\ref{fig:pat1} is the evolution of spatial distribution of
$\tilde\chi$. Initial smooth distribution evolves to random one. Up to
$\eta_*$, the amplitude of $\tilde\chi$ grows keeping its initial
distribution. After $\eta_*$, $\tilde\chi$ begins chaotic behavior and
the correlation of the spatial pattern is lost.  Fig.~\ref{fig:spec1}
shows the evolution of power spectrum of $\tilde\chi$. In this case,
larger scale has  greater power of fluctuation. At $\eta=\eta_*\sim
280$, the fluctuation of the largest scale reaches non-linear and the
characteristic scale  $k_*$ appears in the form of the power
spectrum. For $k<k_*$, the initial power index $-3$ changes to be $1$
that corresponds to white noise. The spectrum for $k>k_*$ keeps its
initial slope and a bend at $k=k_*$ appears in the power spectrum. As
time goes on, the value of $k_*$ becomes larger and the spectrum for
$k<k_c$ becomes that of white noise $\mathcal{P}\propto k$ when $k_*$
reaches $k_c$. Thus, although the large scale fluctuation reaches
$O(1)$ and becomes non-linear temporally, its amplitude decreases to
be smaller than unity when the fluctuation with smaller scale reaches
non-linear.

\paragraph{$n=-1$ case}

\begin{figure}[H]
  
  \centering
\includegraphics[width=0.22\linewidth,clip]{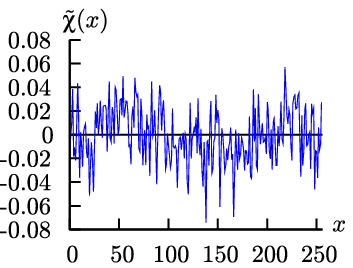}\hspace{0.5cm}
\includegraphics[width=0.22\linewidth,clip]{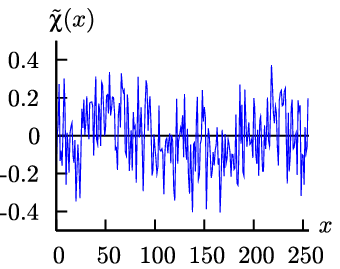}\hspace{0.5cm}
\includegraphics[width=0.22\linewidth,clip]{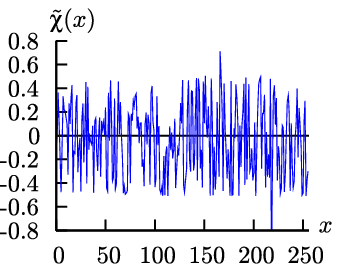}\hspace{0.5cm}
\includegraphics[width=0.22\linewidth,clip]{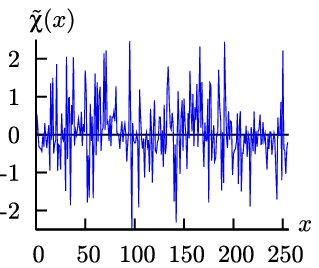}
  \caption{\label{fig:pat2}Evolution of the spatial distribution of $\tilde\chi$ field
    for the initial power index $n+1=0$. From the left to the right
    panel, $\eta=0, 260, 300, 400$.}
\end{figure}

\begin{figure}[H]

  \centering
\includegraphics[width=0.4\linewidth,clip]{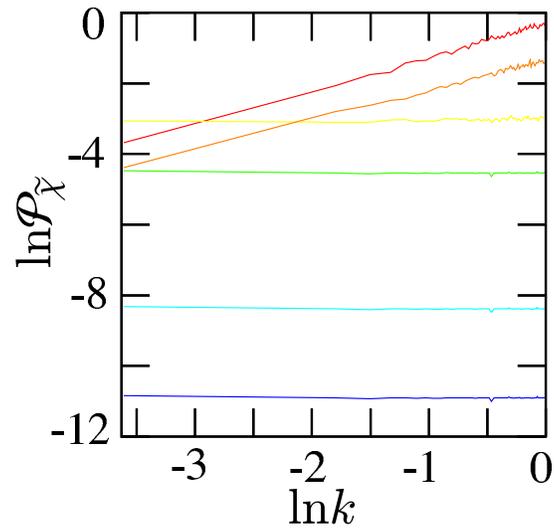}
  \caption{\label{fig:spec2}Evolution of the  power spectrum  of
    $\tilde\chi$ field for the initial power index $n+1=0$ at
    $\eta=0\text{(blue)}, 100\text{(cyan)}, 200\text{(green)},
    300\text{(yellow)}, 400\text{(orange)},500\text{(red)}$.}
\end{figure}

In this case, the initial power spectrum of the $\tilde\chi$ field is
scale invariant. Fig.~\ref{fig:pat2} shows the evolution of spatial
distribution of $\tilde\chi$ and Fig.~\ref{fig:spec2} shows the
evolution of power spectrum of $\tilde\chi$. Before $\eta_*$, the
spectrum evolves keeping its initial power index $0$ and all scales
goes to non-linear simultaneously at $\eta=\eta_*$. Then the power
index changes to be $1$ for all $k$. As the same as $n=-3$ case, the
power of the large scale fluctuation is suppressed after  the
amplitude of the fluctuation reaches non-linear.

\paragraph{$n=0$ case}

\begin{figure}[H]
  \centering
 \includegraphics[width=0.22\linewidth,clip]{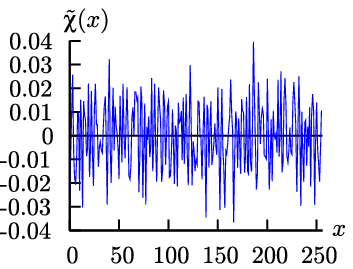}\hspace{0.5cm}
 \includegraphics[width=0.22\linewidth,clip]{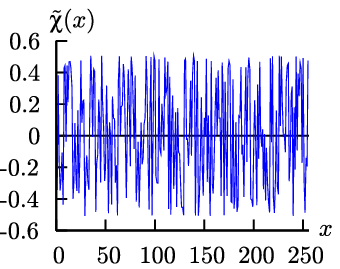}\hspace{0.5cm}
 \includegraphics[width=0.22\linewidth,clip]{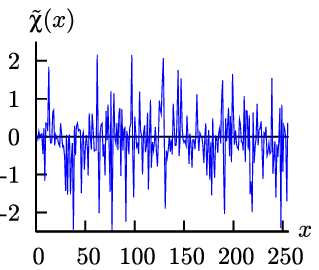}\hspace{0.5cm}
 \includegraphics[width=0.22\linewidth,clip]{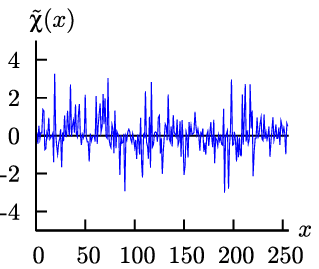}
  \caption{\label{fig:pat3}Evolution of the spatial distribution of $\tilde\chi$ field
    for the initial power index $n+1=1$. From the left to the right
    panel, $\eta=0, 100, 200, 300$.}
\end{figure}

\begin{figure}[H]
  \centering
 \includegraphics[width=0.4\linewidth,clip]{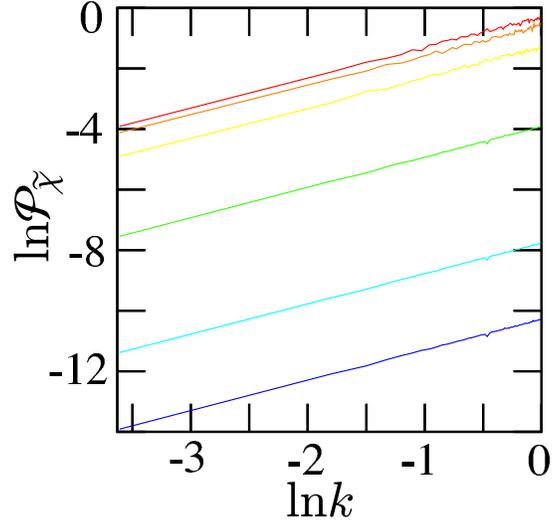}
  \caption{\label{fig:spec3}The power spectrum  of $\tilde\chi$ field for the initial
    power index $n+1=1$ at $\eta=0\text{(blue)},
    100\text{(cyan)}, 200\text{(green)},
    300\text{(yellow)}, 400\text{(orange)},500\text{(red)}$.}
\end{figure}

In this case, the initial spectrum is same as that of white noise and
the small scale reaches non-linear first. Fig.~\ref{fig:pat3} shows
the evolution of spatial distribution of $\tilde\chi$ and
Fig.~\ref{fig:spec3} shows the evolution of power spectrum of
$\tilde\chi$. As the initial spectrum is same as that of white noise,
the power index does not change during whole evolution.

\paragraph{$n=1$ case} 

\begin{figure}[H]
  \centering
\includegraphics[width=0.22\linewidth,clip]{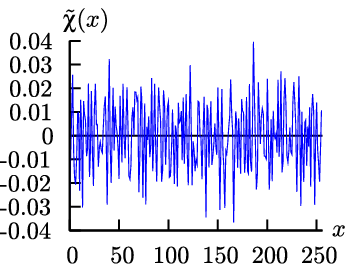}\hspace{0.5cm}
\includegraphics[width=0.22\linewidth,clip]{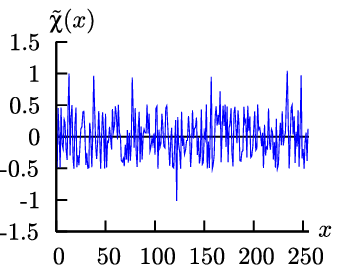}\hspace{0.5cm}
\includegraphics[width=0.22\linewidth,clip]{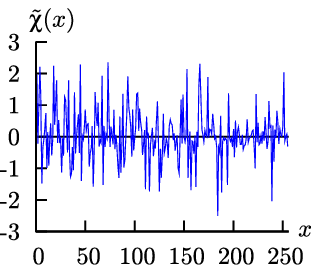}\hspace{0.5cm}
\includegraphics[width=0.22\linewidth,clip]{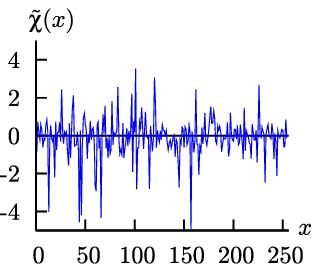} 
 \caption{\label{fig:pat4}Evolution of the spatial distribution of $\tilde\chi$ field
   for the initial power index  $n+1=2$. From the left to the right
   panel, $\eta=0, 260, 300, 400$.}
\end{figure}

\begin{figure}[H]
  \centering
\includegraphics[width=0.4\linewidth,clip]{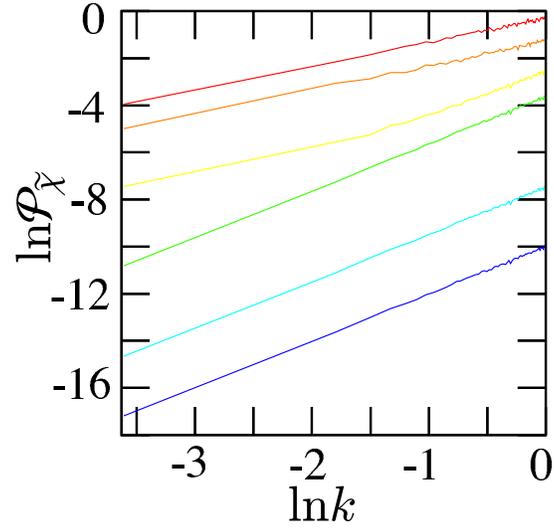}
  \caption{\label{fig:spec4}Evolution of the power spectrum  of
    $\tilde\chi$ field for the initial power index  $n+1=2$ at
    $\eta=0\text{(blue)}, 100\text{(cyan)}, 200\text{(green)},
    300\text{(yellow)}, 400\text{(orange)},500\text{(red)}$.}
\end{figure}

In this case, the initial power index  is larger than that of the
white noise. Fig.~\ref{fig:pat4} shows the evolution of spatial
distribution of $\tilde\chi$ and Fig.~\ref{fig:spec4} shows the
evolution of power spectrum of $\tilde\chi$. After the small scale
reaches non-linear, the initial power index $2$ reduced to be $1$.

\subsection{Cosmological case}
\begin{figure}[H]
  \centering
  \includegraphics[width=0.4\linewidth,clip]{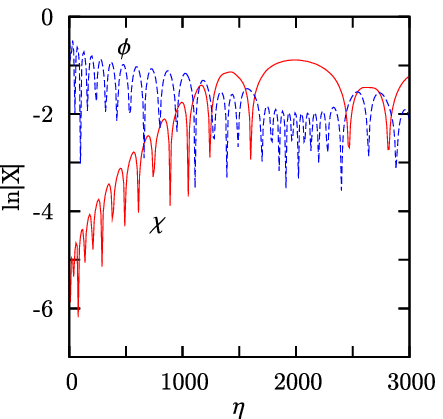}%
 \includegraphics[width=0.4\linewidth,clip]{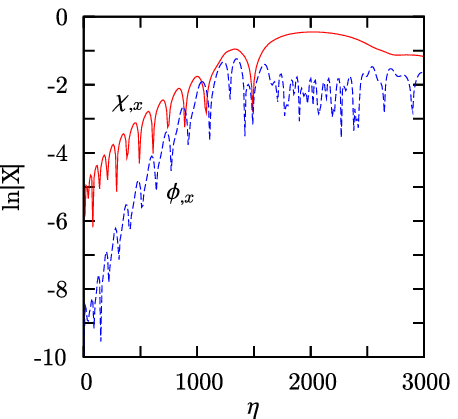}%
  \caption{In the left panel, evolution of scalar fields at a specific
    spatial point is shown. In the right  panel, evolution of the spatial
    derivative of the scalar fields is shown.}
  \label{fig:ge-local}
\end{figure}

\begin{figure}[H]
  \centering
  \includegraphics[width=0.4\linewidth,clip]{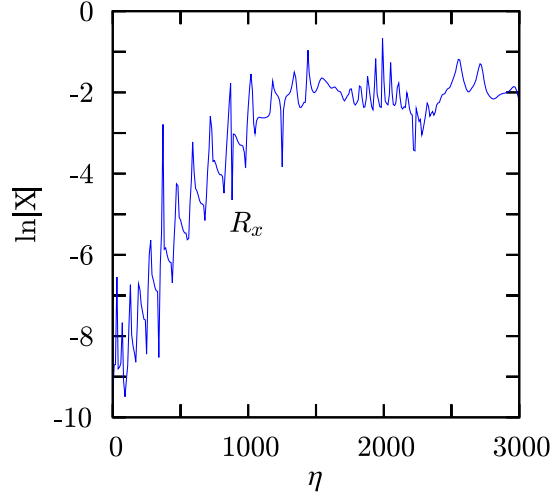}
  \caption{Evolution of the curvature variable 
    $\mathcal{R}_i$ at a specific spatial point.}
  \label{fig:Ri-local}
\end{figure}

We next solve Eqs~\eqref{eq:sync1}, \eqref{eq:sync2},
\eqref{eq:sync3}, and investigate the evolution of the long wavelength
metric fluctuations.   The behavior of variables at a specific spatial
point is shown in Fig.~\ref{fig:ge-local}  and Fig.~\ref{fig:Ri-local}.
\noindent
The characteristic cosmic time when the system becomes non-liner is
given by
$$
 t_*=-\frac{1}{2H_0}+\frac{H_0}{2}\eta_*^2\sim 1460.
$$
Until this time, $\chi_{,x}$ grows exponentially in conformal time and
$\phi_{,x}$ also grows exponentially in accord with the growth of
$\chi_{,x}$ with larger growth rate than that of $\chi_{,x}$. We must
take care that the amplitude of the spatial derivative of the scalar
fields does not reach unity at $t\sim t_*$. This is due to the suppression of
the cosmic expansion but the non-linear regime begins at $t_*$. At
$t\sim t_*$,  $\chi_{,x}\sim\phi_{,x}$ and the back reaction of $\chi$ on
the evolution of $\phi$ begins to be significant and the growth of the
fluctuation caused by the parametric resonance stops. The evolution of
metric variables $H, \al, \mathcal{R}_i$ is same as the evolution of
$\chi_{,x}$ and they grow until $t\sim t_*$ when the back reaction
cannot be neglected.

\begin{figure}[H]
  \centering
  \includegraphics[width=0.4\linewidth,clip]{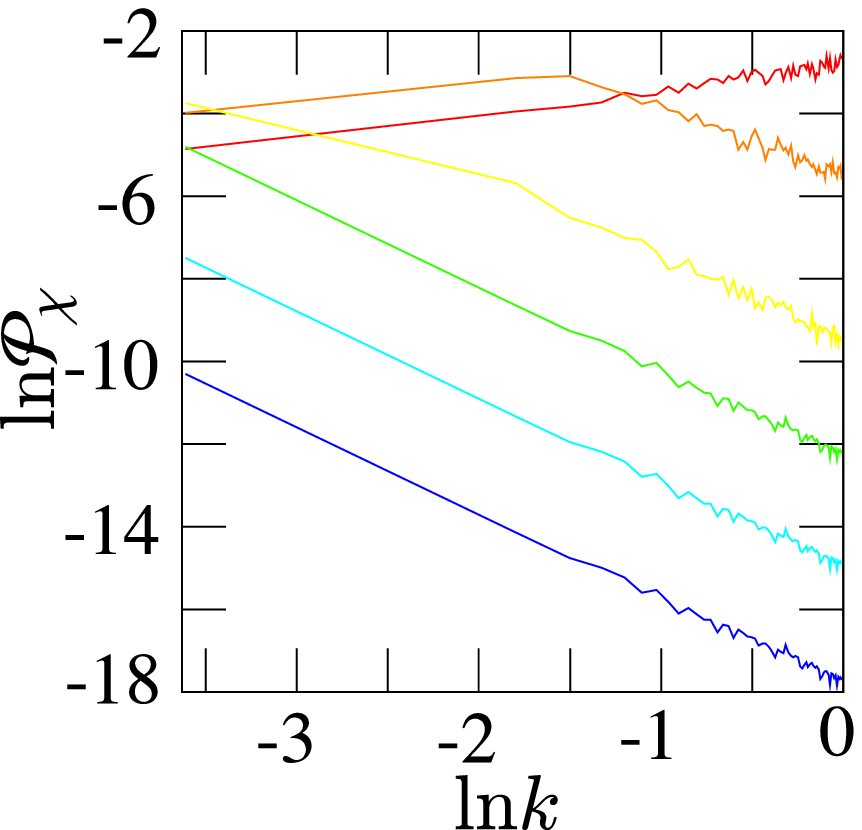}%
  \hspace{1cm}
  \includegraphics[width=0.4\linewidth,clip]{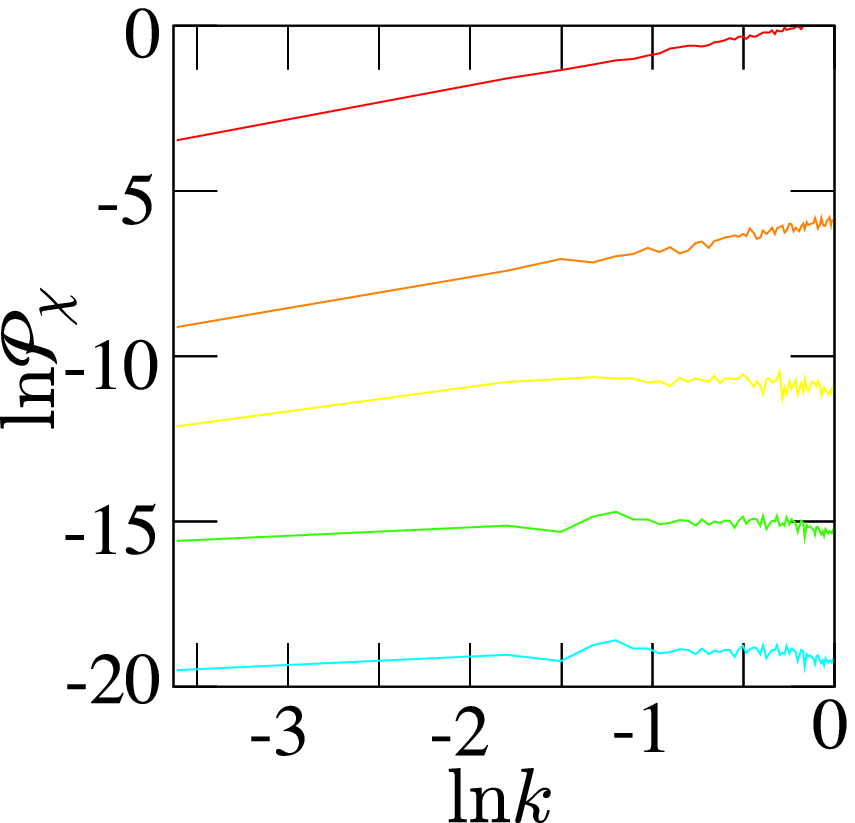}%
   \caption{\label{fig:gspec1}Evolution of the power spectrums for the
     initial power index $n+1=-2$  case. The left panel is
     $\mathcal{P}_{\chi}$ and the right panel is
     $\mathcal{P}_{\mathcal{R}_i}$. Time is $t=0$(blue), 400(cyan),
     800(green), 1200(yellow), 1600(orange), 2000(red). }
\end{figure}

Evolution of the power spectrums of $\chi$ and $\mathcal{R}_i$ depends
on the initial power index of the $\chi$ field.  The evolution of the
power spectrum of $\chi$ is qualitatively same as the flat space
case. For $n=-3$ (Fig.~\ref{fig:gspec1}), until $t\sim t_*$, the
spectrum of $\mathcal{R}_i$ keeps its initial flat shape $k^{0}$.  At
$t\sim t_*$, the large scale fluctuation of $\chi$ field goes to be
non-linear and the bend of the spectrum of $\mathcal{R}_i$ at $k=k_*$
appears. The spectrum of $\mathcal{R}_i$ becomes $k$ for $k<k_*$.  The
amplitude of the curvature variable $\mathcal{R}_i$ stays less than
unity during whole evolution.

\begin{figure}[H]
  \centering
  \includegraphics[width=0.4\linewidth,clip]{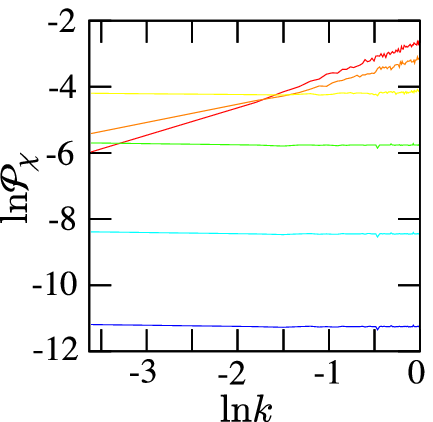}%
  \hspace{1cm}
  \includegraphics[width=0.4\linewidth,clip]{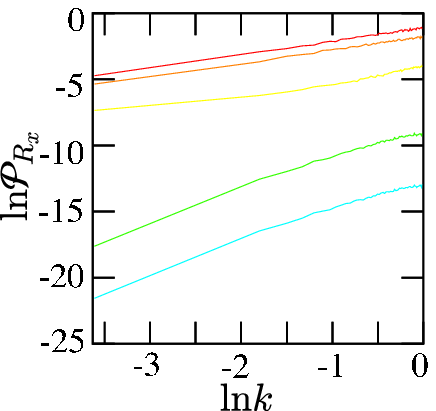}%
 \caption{\label{fig:gspec2}Evolution of the power spectrums for the
   initial power index $n+1=0$ case. The left panel is
   $\mathcal{P}_{\chi}$ and the right panel is
   $\mathcal{P}_{\mathcal{R}_i}$. Time is
   $t=0\text{(blue)},400\text{(cyan)},800\text{(green)},
   1200\text{(yellow)},1600\text{(orange)},2000\text{(red)}$.}
\end{figure}

\begin{figure}[H]
  \centering
  \includegraphics[width=0.4\linewidth,clip]{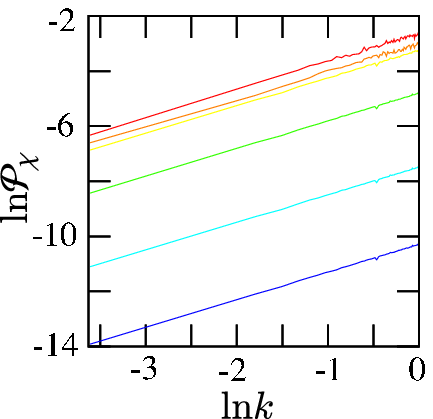}%
\hspace{1cm}
  \includegraphics[width=0.4\linewidth,clip]{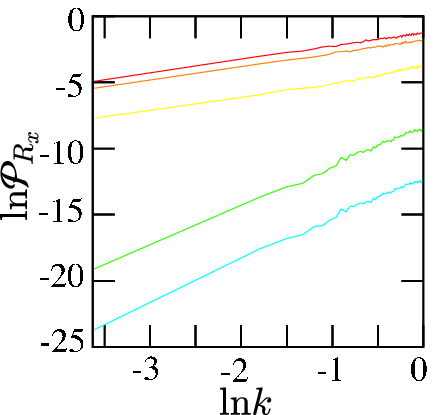}%
 \caption{\label{fig:gspec3}Evolution of the power spectrums for the initial power index
   $n+1=1$ case. The left panel is $\mathcal{P}_{\chi}$ and the right
   panel is  $\mathcal{P}_{\mathcal{R}_i}$. Time is
   $t=0\text{(blue)},400\text{(cyan)},800\text{(green)},
1200\text{(yellow)},1600\text{(orange)},2000\text{(red)}$.}
\end{figure}

\begin{figure}[H]
  \centering
  \includegraphics[width=0.4\linewidth,clip]{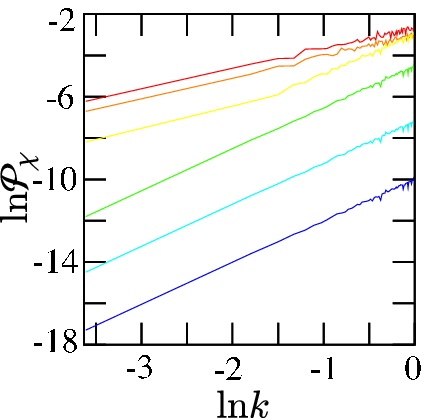}%
  \hspace{1cm}
  \includegraphics[width=0.4\linewidth,clip]{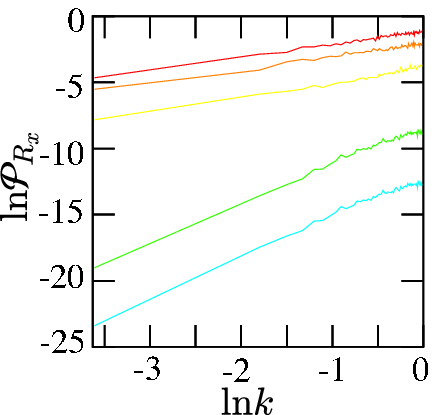}%
 \caption{\label{fig:gspec4}Evolution of the power spectrums for the
   initial power index $n+1=2$ case. The left panel is
   $\mathcal{P}_{\chi}$ and the right panel is
   $\mathcal{P}_{\mathcal{R}_i}$. Time is
   $t=0\text{(blue)},400\text{(cyan)},800\text{(green)},
   1200\text{(yellow)},1600\text{(orange)},2000\text{(red)}$.}
\end{figure}

For $n\geq -1$ (Fig.~\ref{fig:gspec2}, Fig.~\ref{fig:gspec3},
Fig.~\ref{fig:gspec4}),  until $t\sim t_*$, the spectrum of
$\mathcal{R}_i$ keeps its initial shape $k^{n+3}$.  At $t\sim t_*$,
the small scale fluctuation of $\chi$ field  goes to be non-linear and
after this time, the spectrum of $\mathcal{R}_i$ becomes $k$ for all
wave number. The amplitude of the curvature variable $\mathcal{R}_i$
stays less than unity during whole evolution.

\section{Summary and discussion}

In this paper, we investigated the evolution  of long wavelength
fluctuations during preheating after inflation. By using the separate
universe approach, we obtained  the evolution of the power spectrum of
long wavelength fluctuations numerically. During the linear stage of
the evolution, the fluctuation of the massless field grows
exponentially in time by the parametric amplification, but the power
spectrum keeps its initial shape. When  the fluctuation of the
massless field becomes non-linear, the amplification of fluctuations
stops by the back reaction effect of massless field on the inflaton
field.  After this time, the system enters the chaotic non-linear
stage and  the shape of the spectrum changes. For large scale mode,
the amplitude of the fluctuation is suppressed because the statistical
nature of the large scale mode becomes that of white noise after the
non-linear evolution. The evolution of the curvature variable that has
an important role in the cosmological scenario is qualitatively same
as the evolution of the scalar field fluctuations. Hence we do not
expect the significant effect of the parametric amplification  during
preheating on the evolution of the large scale fluctuations of metric
variables.  On the other hand, for small scale fluctuations, the power
grows for larger $k$ and this leads to the possibility of the
formation of the non-linear structures such as primordial black
holes\cite{SuyamaT:PRD71:2005}.

\begin{acknowledgments}
This work was supported in part by a Grant-In-Aid for Scientific
Research of the Ministry of Education, Science, Sports, and Culture of
Japan (11640270).
\end{acknowledgments}

\begin{thebibliography}{10}

\bibitem{TraschenJ:PRD42:1990}
Traschen J and Brandenberger R~H 1990 Particle production during
  out-of-equilibrium phase transitions \emph{Phys. Rev. D} \textbf{42}
  2491--2504

\bibitem{KofmanL:PRL73:1994}
Kofman L, Linde A and Starobinsky A 1994 Reheating after Inflation \emph{Phys.
  Rev. Lett.} \textbf{73} 3195--3198

\bibitem{KodamaH:PTP96:1996}
Kodama H and Hamazaki T 1996 Evolution of Cosmological Perturbations in a Stage
  Dominated by an Oscillatory Scalar Field \emph{Prog. Theor. Phys.}
  \textbf{96} 949--970

\bibitem{NambuY:1997}
Nambu Y and Taruya A 1997 Evolution of Cosmological Perturbation in Reheating
  Phase of the Universe \emph{Prog. Theor. Phys.} \textbf{97} 83--89

\bibitem{TaruyaA:1998}
Taruya A and Nambu Y 1998 Cosmological Perturbation with two scalar fields in
  Reheating after Inflation \emph{Phys. Lett.} \textbf{B428} 37--43

\bibitem{BassetB:PRD62:2000}
Basset B and Viniegra F 2000 Massless metric preheating \emph{Phys. Rev. D}
  \textbf{62} 043507

\bibitem{ZibinJP:PRD63:2001}
Zibin J~P, Brandenberger R~H and Scott D 2001 Backreaction and the parametric
  resonance of cosmological fluctuations \emph{Phys. Rev. D} \textbf{63} 043511

\bibitem{WandsD:2000}
Wands D, Malik K~A, Lyth D~H and Liddle A~R 2000 New approach to the evolution
  of cosmological perturbations on large scales \emph{Phys. Rev. D} \textbf{62}
  043527

\bibitem{NambuY:PRD71:2005}
Nambu Y 2005 The separate universe and the backreaction of long wavelength
  fluctuations \emph{Phys. Rev. D} \textbf{71} 084016:1--5

\bibitem{TanakaT:2003}
Tanaka T and Basset B 2003 Application of the Separate Universe Approach to
  Preheating \emph{astro-ph/0302544}

\bibitem{SalopekDS:1992}
Salopek D~S and Stewart J~M 1992 Hamilton-Jacobi theory for General Relativity
  with Matter Fields \emph{Class. Quantum Grav.} \textbf{9} 1943--1967

\bibitem{RigopoulosGI:2003}
Rigopoulos G~I and Shellard E~P~S 2003 Separate Universe Approach and the
  Evolution of Nonlinear Superhorizon Cosmological Perturbations \emph{Phys.
  Rev. D} \textbf{68} 123518:1--8

\bibitem{PBGreeneLKofmanAL:PRD56:1997}
Greene P~B, Kofman L, Linde A and Starobinsky A~A 1997 Structure of resonance
  in preheating after inflation \emph{Phys. Rev. D} \textbf{56} 6175--6192

\bibitem{SuyamaT:PRD71:2005}
Suyama T, Tanaka T, Basset B and Kudoh H 2005 Are black holes overproduced
  during preheating? \emph{Phys. Rev. D} \textbf{71} 063507

\end{thebibliography}

\end{document}